\documentclass[ArXiv,AXISWP,accept,moreauthors,pdftex,10pt,letterpaper]{Definitions/axis} 
\usepackage{tcolorbox, amssymb}
\usepackage{floatrow, bm}

\firstpage{1} 
\pubvolume{xx}
\issuenum{1}
\articlenumber{5}
\pubyear{2023}
\copyrightyear{2023}

\Title{AGN in overdense environments at high-$z$ with AXIS     }

\Author{Fabio Vito,$^{1}$ Paolo Tozzi, $^{2}$ Roberto Gilli $^{1}$, Stefano Marchesi$^{1,3,4}$, Nico Cappelluti,$^{5}$ Adi Foord$^{6}$}

\AuthorNames{Fabio Vito, Paolo Tozzi, Roberto Gilli, Stefano Marchesi, Nico Cappelluti, Adi Foord}

\address{%
$^{1}$ \quad INAF-OAS, via Piero Gobetti 93/3, 40129 Bologna, Italy\\
$^{2}$ \quad INAF-OAA, Largo Enrico Fermi 5, 50125 Firenze, Italy\\
$^{3}$ \quad Department of Physics and Astronomy, Clemson University, Kinard Laboratory of Physics, 140 Delta Epsilon Ct., Clemson, SC 29634, USA \\
$^{4}$ \quad Dipartimento di Fisica e Astronomia, Università degli Studi di Bologna, via Gobetti 93/2, 40129 Bologna, Italy\\
$^{5}$ \quad Department of Physics, University of Miami, Coral Gables, FL 33124, USA\\
$^{6}$ \quad Department of Physics, University of Maryland Baltimore County, 1000 Hilltop Cir, Baltimore, MD 21250, USA
}

\abstract{
Overdense regions at high redshift ($z \gtrsim 2$) are perfect laboratories to study
the relations between environment and SMBH growth, and the AGN feedback processes on the surrounding galaxies and diffuse gas. In this white paper, we discuss how AXIS will 1) constrain the AGN incidence in protoclusters, as a function of parameters such as redshift, overdensity, mass of the structure; 2) search for low-luminosity and obscured AGN in the satellite galaxies of luminous QSOs at $z>6$, exploiting the large galaxy density around such biased objects; 3) probe the AGN feedback on the proto-ICM via the measurement of the AGN contribution to the gas ionization and excitation, and the detection of extended X-ray emission from the ionized gas and from radio jets; 4) discover new large-scale structures in the wide and deep AXIS surveys as spikes in the redshift distribution of X-ray sources. These goals can be achieved only with an X-ray mission with the capabilities of AXIS, ensuring a strong synergy with current and future state-of-the-art facilities in other wavelengths.
\emph{This White Paper is part of a series commissioned for the AXIS Probe Concept Mission; additional AXIS White Papers can be found at the  \href{http://axis.astro.umd.edu/}{AXIS website} with a mission overview \href{https://arxiv.org/abs/2311.00780}{here}}.
}


\begin{document}
\tableofcontents
\listoffigures


\section{Introduction}
The large-scale environment has a major influence on galaxy evolution,
which is strongly accelerated in high-redshift dense regions (e.g., \cite{Kauffmann96, Delucia06, Zeimann12}).    The high gas consumption, the strong nuclear feedback, and the high merger rate for member galaxies, cause
massive, red, and passive spheroids to dominate the galaxy population in the cores of the most massive (i.e., $M_{vir}>10^{14}\,M_\odot$)
gravitationally bound and virialized objects in the local universe, i.e., galaxy clusters (e.g., \cite{Smith08, Willis20}). Most of the baryonic mass of galaxy clusters consists of a hot ($>10^7$ K), tenuous, and optically thin plasma, the intracluster medium (ICM; e.g., \cite{Gonzalez13}), which is partially enriched 
in metals by past star-formation events in the cluster galaxies and that has been found to be a source of X-ray emission up to redshift $z=2-2.5$ (\cite{Tozzi22b,WangT16}). Feedback from accreting SMBHs is fundamental to keep the ICM hot, and prevent the 
formation of massive cooling flows infalling into the cluster galaxies (e.g., \cite{Birzan17}), particularly
into the central brightest galaxy (BCG). 
In fact, bright radio galaxies often reside in the cluster centers, 
and are powered by the most massive ($>10^9\,\mathrm{M_\odot}$) SMBHs of the local Universe (e.g., \cite{Bassini19}), 
implying that most of their mass, as those of their host galaxies, has been built up at earlier cosmic epochs.

The progenitors of galaxy clusters, i.e., high-redshift ($z\gtrsim2$) protoclusters 
(e.g., \cite{Overzier16})  are unique laboratories to probe the early phases of galaxy
and SMBH growth, AGN feedback, and ICM formation, which shaped the observed properties of their present-day descendants. The number of known protoclusters is still relatively low, 
but it has been constantly increasing in the last $\sim10-20$ yrs thanks to the
applications of several selection techniques, including wide-field 
spectroscopic and photometric surveys (e.g., \cite{Cucciati14, Toshikawa16}), 
narrow-band imaging (\cite{Ouchi05, Higuchi19}), 
sub-mm observations (e.g., \cite{Oteo18, Hill20}), and the use of objects expected to trace highly
biased regions of the universe, such as powerful radio galaxies
(e.g., \cite{Pentericci00, Venemans02, Gilli19, Brienza23}), QSOs up to $z=6.3$ (e.g., \cite{Mignoli20}), and extended Ly$\alpha$ emission over tens-to-hundreds kpc scales (e.g., \cite{Cantalupo14,
ArrigoniBattaia22}). As a consequence, the term {\sl protocluster} is used to designate a large variety 
of large scale overdensity across different epochs, with vastly different properties.

\begin{figure}
\centering
\includegraphics[width=\textwidth]{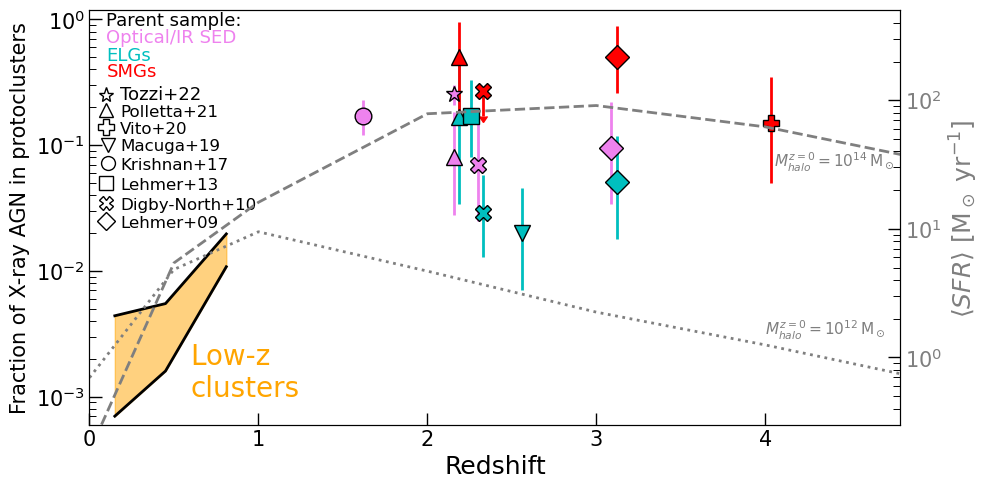}
\caption[X-ray AGN fraction in protoclusters.]{Fraction of X-ray selected AGN among member galaxies of eight massive protoclusters
collected by \cite{Tozzi22a} (see different symbols in the figure). Results associated with different selection methods
for the parent galaxy population (i.e., optical/IR SED, Emission-Line Galaxies, Sub-mm Galaxies)
are reported with different colours. For three protoclusters, marked with upward-pointing triangle,
cross, and diamond symbols, all of the three selection methods have been employed separately,
showing their impact on the estimated AGN fraction. We used the published values or we computed
the fractions based on the published information. The gold stripe marks the AGN fraction in low-
redshift galaxy clusters (\cite{Martini13}). The dashed and dotted gray lines are the average SFR of galaxies in dark-matter halos with masses at $z=0$ typical of cluster and field environments, respectively, as computed by \cite{Behroozi13}.}
\end{figure}

X-ray observations are fundamental to identify AGN among protocluster galaxies, even in the presence of heavy obscuration, which often characterizes AGN in dense and gas-rich regions (e.g., \cite{Gilli19,Vito20}), investigate their physical properties, and probe AGN feedback in those structures. However, our knowledge on the AGN population in protoclusters is still limited to a small number of structures with sensitive X-ray coverage obtained via time-expensive observations mainly with Chandra. The unique combination of large effective area, sharp angular resolution, and low background level of AXIS (\cite{2023_AXIS_Overview}) is an unprecedented tool for investigating SMBH accretion in a considerable number of high-redshift protoclusters with relatively short exposure times. The results will allow us to unveil the link between dense environment and SMBH growth, and the effects of AGN feedback on star formation and proto-ICM at $z\gtrsim2$. In this respect, the AXIS capabilities will ensure a strong synergy with current and future state-of-the-art astronomical facilities, such as the Vera C. Rubin
Observatory, the Roman Observatory, SUBARU/PSF, EUCLID, and the South Pole Telescope, which will discover and characterize thousands of high-redshift protoclusters, but cannot unveil the complete AGN population in those structures.


\section{Enhanced SMBH growth in protoclusters}
Semi-analytic models and numerical simulations find that star-formation is enhanced in the high-redshift regions that will collapse in present-day clusters (see gray lines in Fig. 1), such that
protoclusters contribute to a significant
fraction of the cosmic star-formation rate density at a given epoch (20--50\%; e.g., \cite{Chiang17}). As for galaxy growth, the availability of huge gas reservoirs and
the high rate of galaxy interactions in protoclusters (e.g., \cite{Hine16, Oteo18}) are thought
to promote SMBH accretion, possibly obscured by large gas column densities (e.g., \cite{Assef15, Vito20}), by providing the fuel and the mechanisms to drive it into the
SMBH potential wells at the centers of the host galaxies (e.g., \cite{Hopkins08, Sabater15,
Satyapal17}), resulting in a higher AGN duty cycle than in the field environment and in low-redshift clusters.


Current Chandra observations aiming to study the AGN population in protoclusters are able to identify X-ray sources down to $L_X=10^{43}-10^{44}\,\mathrm{erg\,s^{-1}}$
in $z\approx2-4$ structures, even in the presence of heavy obscuration, but require exposure times of  $\approx$100--700 ks (e.g., \cite{Lehmer09, DigbyNorth10, Macuga19, Vito20, Tozzi22a}). 
Generally, they find significantly higher fractions of AGN among the structure members than in local clusters (Fig. 1), and indications of an enhancement with respect to the field environment at a similar redshift.  
The high incidence of AGN activity in protoclusters might produce strong AGN feedback on the host and surrounding galaxies, as recently found in a $z = 1.7$ protocluster (\cite{Gilli19}), and is expected to eventually suppress the gas infall, star-formation, and SMBH growth itself in the structures, thus shaping the observational properties of local galaxy clusters.

However, the interpretation of these results is hindered by their large scatter, due to the limited sample of high-redshift protoclusters observed in the X-ray band and the inhomogeneous sensitivities of the observations. In fact, the required costly exposure times prevent large samples from being assembled with sensitive coverage.  Moreover, the different
techniques used to identify protocluster galaxies introduce strong selection effects, as different galaxy populations (e.g., sub-mm galaxies and emission-line galaxies) are characterized by intrinsically different AGN content (Fig. 1).  
The main effect of these issues is that a coherent measurement of AGN
incidence in protoclusters is still lacking, strongly affecting our understanding of the influence of environment on SMBH growth. 

AXIS will identify AGN in $z=2-4$ protoclusters with luminosities down to $L_X=10^{43}\,\mathrm{erg\,s^{-1}}$ with exposure times of 5--20 ks, i.e., more than one order of magnitude shorter pointings than the currently available Chandra observations. Therefore, it is foreseeable that AXIS will observe tens to hundreds of protoclusters during its lifespan, which might be collected from the objects discovered by facilities such as Rubin and Euclid. The large sample of high-redshift structures with deep AXIS coverage will allow us to perform a statistical study of the incidence of AGN in protoclusters as a function of cosmic epoch, structure overdensity and mass, selection method of the parent population of galaxies, and compare the results with blank fields, providing us with the first complete picture of the environmental effects on SMBH growth at high redshift.

\section{AGN in overdense regions of the early Universe}
Theory and numerical simulations predict that the large-scale environment has a fundamental role not only in the growth of SMBHs at high redshift, but also on their formation soon after the Big Bang (e.g., \cite{Costa14, Habouzit19}). In fact, both the formation of the BH seeds and the efficient SMBH accretion are highly favoured in the peaks of the dark-matter distribution at $z \approx 10 - 30$. As a direct consequence, luminous and massive QSOs in the early Universe ($z\gtrsim6$) are expected to reside in regions characterized by galaxy counts in excess to the average field (Fig. 2; e.g., \cite{Costa14, Barai18, Habouzit19}), as also confirmed for some objects with ground-based optical/IR facilities (e.g., \cite{Ota18, Mignoli20}) and, more recently, with JWST (e.g., \cite{Kashino23, Wang23}).

Due to the large density of satellite galaxies, the regions around $z\gtrsim6$ QSOs are also those with the highest chances to host more elusive AGN classes, such as low-to-moderate luminosity and obscured AGN. These objects constitutes the vast majority of AGN at high redshift (e.g., \cite{Vito18a, Gilli22}) and carry important information on physics of SMBH formation and early growth, but only a few of them have been identified in deep optical/IR observations (e.g., \cite{Fujimoto22, Endsley23, Maiolino23}). About 50 $z>6$ QSOs have been currently observed with sensitive (tens-to-hundreds ks) X-ray exposures with Chandra and XMM-Newton (\cite{Lusso23}), but none of the known satellite galaxies have been detected with high confidence (e.g., \cite{Vito19, Connor20, Vito21}). This can be due to either a low incidence of nuclear accretion in such galaxies, due to, e.g., low BH occupation fraction or AGN duty cycle, or the limited sensitivity of the available X-ray datasets.

 AXIS will push by an order of magnitude the depth of observations of $z>6$ QSOs, allowing us not only to study the accretion physics of the central object, but also to search for evidence of low-rate and/or obscured accretion in the surrounding galaxies. A $\approx50$ ks AXIS pointing of a $z>6$ QSO can detect from a few to a few tens of satellite AGN. These numbers are computed by
extrapolating state-of-the-art AGN X-ray luminosity functions (e.g., \cite{Vito18a}) from $z\approx4$ to $z>6$, and renormalizing them by the best measurement of galaxy overdensity around a $z>6$ QSO (\cite{Mignoli20}), and are consistent with the expectations from the numerical simulations of \cite{Costa14}. As a comparison, 0--1 satellite AGN are expected to be detected with Chandra in a 500 ks observation.

\begin{figure}
\centering
\includegraphics[width=\textwidth]{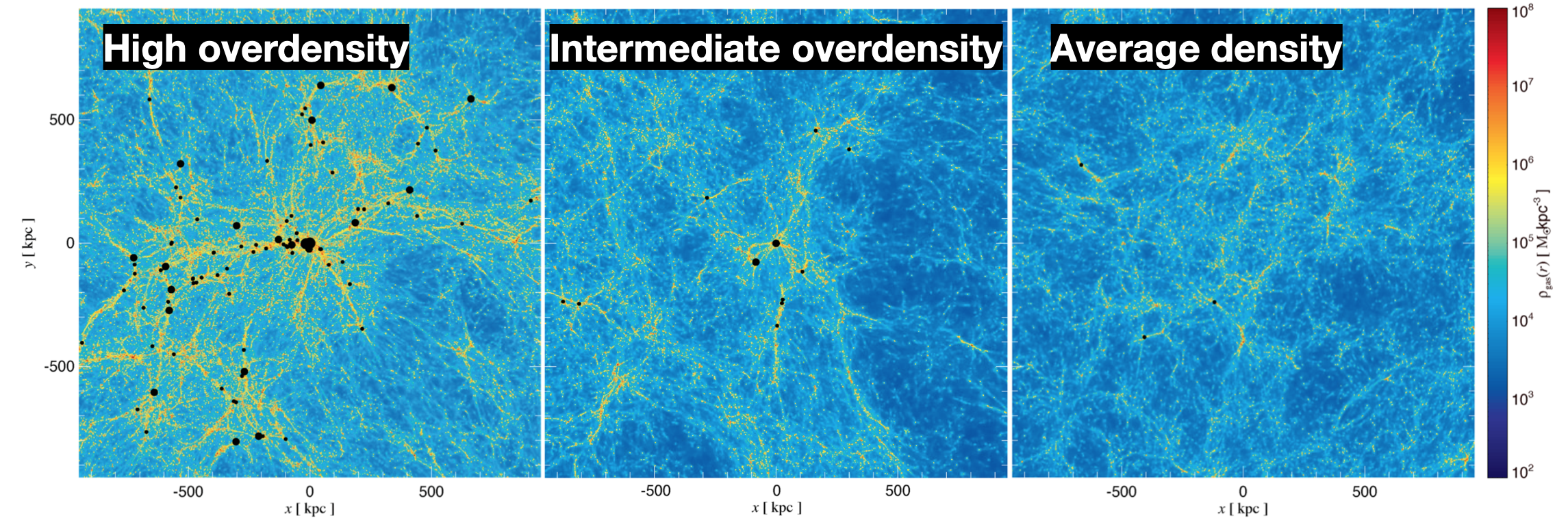}
\caption[Large-scale environment of $z \approx 6$ SMBHs in numerical simulations.]{Gas density maps ($2\times2$ pMpc) for different environments from $z \approx 6$ cosmological simulations. The presence of SMBHs (black dots, with sizes proportional to their masses in the range $10^6-10^9\,\mathrm{M_\odot}$) 
is greatly favoured in overdense environment.
Adapted from \cite{Costa14}.
} 
\end{figure}

\section{AGN feedback on the diffuse gas }

    The   gas infalling from the forming cosmic web is what fuels both star-formation and SMBH growth in protocluster members, and is in turn affected by AGN. One of the most striking evidence for the presence of such gas is represented by Ly$\alpha$ Emission-Line nebulae 
    extending up to hundreds kpc in the central regions of protoclusters  (e.g., \cite{Cantalupo14, Hennawi15, ArrigoniBattaia18}), possibly 
    powered by photoionization due to AGN in those structures (e.g., \cite{Geach09, Overzier13}). X-ray observations are key tools to constrain the AGN contribution to the ionization state of the extended gas reservoirs and thus the effect of AGN feedback. The identification of the complete population of AGN in the structures and the reliable  measurement of their intrinsic X-ray luminosities is required to estimate the rate of ionizing photons emitted from AGN that can power the nebulae (e.g., \cite{Valentino16}).

    Moreover, the ratio between Ly$\alpha$ and X-ray luminosities of the diffuse gas can be used to probe if the nebulae are powered by photoionization or collisional
excitation during the gravitational collapse of the gas (e.g., \cite{Haiman00, Geach09}).  Currently, diffuse X-ray thermal emission from the proto-ICM has been rarely detected in protoclusters (e.g., \cite{Gilli19, Tozzi22b}), and only with long exposures with Chandra. A notable example of this approach is the detection of diffuse X-ray emission from a $z\approx2$ Ly$\alpha$ nebula  with a 140 ks observation with Chandra (Vignali+ in prep).

AGN feedback on the surrounding environment can also be produced by relativistic jets launched from the accreting SMBHs. Measuring the diffuse X-ray emission from radio jets due to inverse Compton scattering onto the photons of the cosmic microwave background and possibly local IR photons allows us to directly constrain the magnetic fields and pressure around the jets and lobes, shedding light on the interactions from small to large scales between the relativistic electrons and the diffuse ICM (e.g., \cite{Carilli22, Anderson22, Brienza23}). 

 The AXIS large effective area at soft energies, one order of magnitude larger than Chandra ACIS-I's area, will allow to routinely detect X-ray emission from  cooling gas in protoclusters,  as well as from relativistic jets launched from radiogalaxies, thus opening a new window onto the physics and mechanisms of AGN feedback, and the formation of the ICM in high-redshift structures.

\begin{figure}

	\floatbox[{\capbeside\thisfloatsetup{capbesideposition={right,top},capbesidewidth=0.4\textwidth}}]{figure}[\FBwidth]
	{\caption[X-ray AGN as tracers of large-scale structures.]{Redshift cumulative distribution (zoomed into the $z=2.3-3.2$ range) of X-ray sources (filled symbols) and all galaxies (empty symbols) in the J1030 field (\cite{Marchesi21,Marchesi23}). The spikes in the X-ray source distribution, as the one indicated by the black arrow, trace the overall distribution of galaxies, and therefore can be used as proxies of large-scale structures and overdensities. From \cite{Marchesi23}.}}
	{\includegraphics[width=0.6\textwidth]{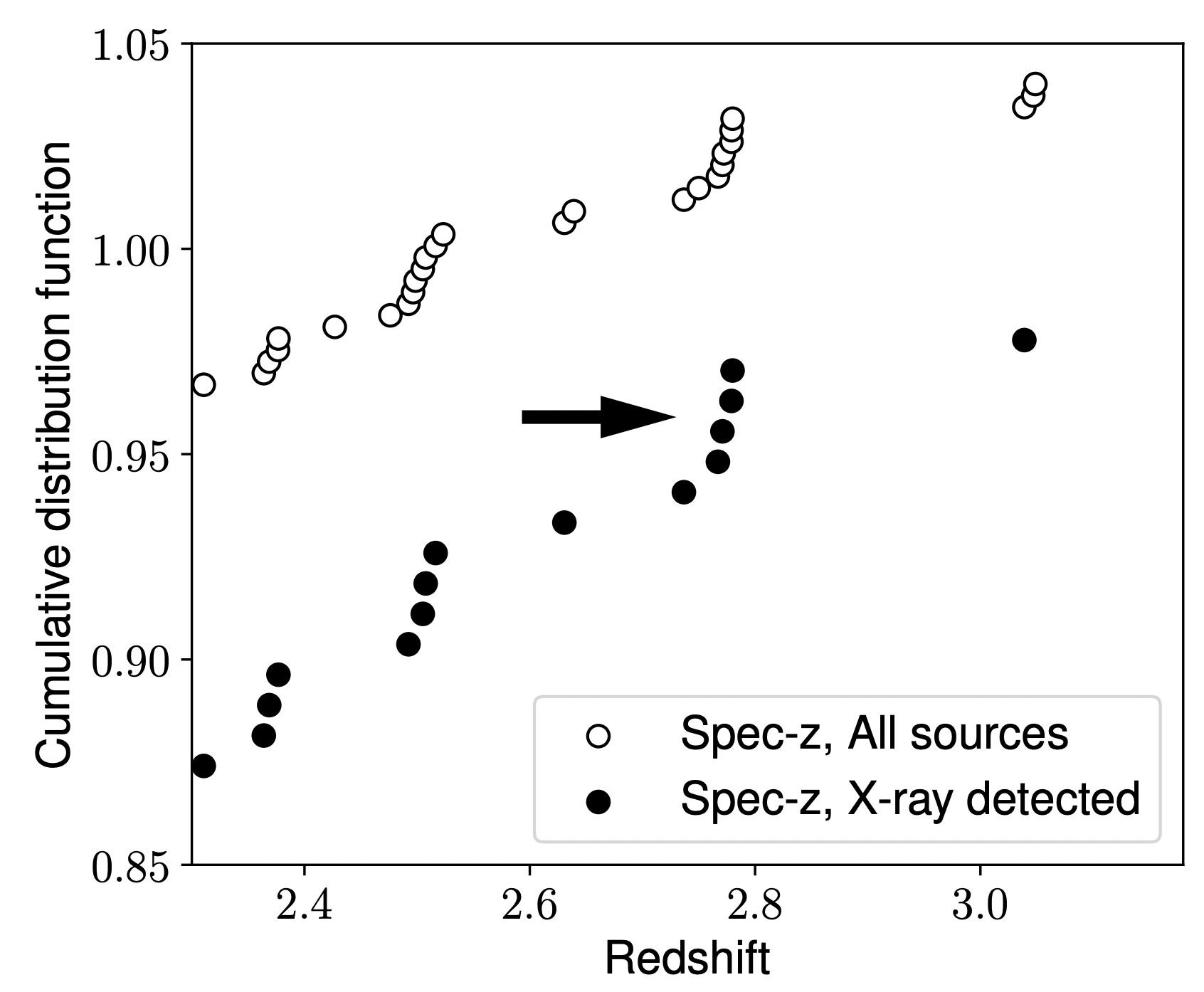}}
\end{figure}

\section{AGN as tracers of protoclusters in the AXIS surveys}

SMBH growth happens preferentially in massive galaxies ($\gtrsim10^{10}\,\mathrm{M_\odot}$; e.g., \cite{Yang17}), which, in turns, are hosted in massive dark-matter haloes and thus are preferentially located in the filaments and nodes of the cosmic web (e.g., \cite{Kuchner22}). Consequently, X-ray selected AGN are thought to be excellent tracers of large-scale structures in X-ray surveys (e.g., \cite{Gilli03, Marchesi21}). Recently, \cite{Marchesi23} demonstrated that even 3--4 X-ray AGN at the same redshift are enough to accurately pinpoint the presence of a galaxy overdensity, and thus possibly a protocluster, due to their low surface density (Fig. 3).
The AXIS "wedding-cake"  extragalactic survey will detect $\approx3\times10^4$ AGN over $\gtrsim2.5\,\mathrm{deg^{-2}}$ (Cappelluti et al. in this series of WPs). Their redshift distributions will pintpoint the presence and locations of likely tens of  new high-redshift structures, among which protoclusters, up to high redshift. Under the reasonable assumption that the AXIS survey will be conveniently performed on regions with deep multi-wavelength coverage, all the required observational datasets will be immediately available to investigate the environmental effect on SMBH growth and evidence for AGN feedback in the discovered protoclusters.

\section{Conclusions}
Protoclusters will be among the most studied astrophysical objects in the next decades, especially in light of future facilities such as Euclid, the Vera Rubin Observatory, the Roman Space Telescope, and the Square Kilometre Array.
An X-ray telescope with sharp angular resolution and large effective area, especially in the soft band, is required to investigate the effects of dense environment on SMBH growth and the resulting AGN feedback, ultimately probing the physical mechanisms that shape the observational properties of local galaxy clusters. 
We discussed how  AXIS will obtain transformational results in some research areas related to the AGN population in overdense regions at $z\gtrsim2$:

\begin{itemize}
    \item it will measure the AGN incidence in a large sample of protocluster, as a function of redshift, overdensity, and mass of the structure, controlling for the selection of the parent population of galaxies. The comparison of the outcomes with those obtained in blank fields will result in an unprecendented view of the impact of overdense and gas-rich environment on BH growth.
    \item It will probe the effect of AGN feedback on the diffuse gas in overdense regions at high redshift, by detecting diffuse X-ray emission from the infalling gas and radio jets.
    \item It will allow us to search for elusive populations of AGN in the early Universe, by detecting X-ray sources hosted in satellite galaxies of $z>6$ QSOs, exploiting the large galaxies densities around such biased objects.
    \item As one of their legacy results, the AXIS surveys will reveal the presence and locations of tens of galaxy overdensities over most of the cosmic time as spikes of the redshift distribution of X-ray selected AGN. 
\end{itemize}

These results can be hardly obtained in other electro-magnetic bands, ensuring a great synergy between AXIS and state-of-the-art multi-wavelength facilities.

\acknowledgments{We kindly acknowledge the AXIS team for their outstanding scientific and technical work over the past year. This work is the result of several months of discussion in the AXIS-AGN SWG.}

\externalbibliography{yes}
\bibliography{biblio}

\end{document}